\documentstyle[12pt]{article}
\advance \topmargin by -\headheight
\advance \topmargin by -\headsep

\evensidemargin \oddsidemargin
\begin{document}

\topmargin -.6in

%
\def\rf#1{(\ref{eq:#1})}
\def\lab#1{\label{eq:#1}}
\def\nonu{\nonumber}
\def\br{\begin{eqnarray}}
\def\er{\end{eqnarray}}
\def\be{\begin{equation}}
\def\ee{\end{equation}}
\def\eq{\!\!\!\! &=& \!\!\!\! }
\def\foot#1{\footnotemark\footnotetext{#1}}
\def\lb{\lbrack}
\def\rb{\rbrack}
\def\llangle{\left\langle}
\def\rrangle{\right\rangle}
\def\blangle{\Bigl\langle}
\def\brangle{\Bigr\rangle}
\def\llbrack{\left\lbrack}
\def\rrbrack{\right\rbrack}
\def\lcurl{\left\{}
\def\rcurl{\right\}}
\def\({\left(}
\def\){\right)}
\newcommand{\nit}{\noindent}
\newcommand{\ct}[1]{\cite{#1}}
\newcommand{\bi}[1]{\bibitem{#1}}
\def\lskip{\vskip\baselineskip\vskip-\parskip\noindent}
\relax

\def\tr{\mathop{\rm tr}}
\def\Tr{\mathop{\rm Tr}}
\def\v{\vert}
\def\bv{\bigm\vert}
\def\Bgv{\;\Bigg\vert}
\def\bgv{\bigg\vert}
\newcommand\partder[2]{{{\partial {#1}}\over{\partial {#2}}}}
\newcommand\funcder[2]{{{\delta {#1}}\over{\delta {#2}}}}
\newcommand\Bil[2]{\Bigl\langle {#1} \Bigg\vert {#2} \Bigr\rangle}  
\newcommand\bil[2]{\left\langle {#1} \bigg\vert {#2} \right\rangle} 
\newcommand\me[2]{\left\langle {#1}\right|\left. {#2} \right\rangle} 
\newcommand\sbr[2]{\left\lbrack\,{#1}\, ,\,{#2}\,\right\rbrack}
\newcommand\pbr[2]{\{\,{#1}\, ,\,{#2}\,\}}
\newcommand\pbbr[2]{\lcurl\,{#1}\, ,\,{#2}\,\rcurl}
%
\def\a{\alpha}
\def\b{\beta}
\def\dc{{\cal D}}
\def\d{\delta}
\def\D{\Delta}
\def\eps{\epsilon}
\def\vareps{\varepsilon}
\def\g{\gamma}
\def\G{\Gamma}
\def\grad{\nabla}
\def\h{{1\over 2}}
\def\l{\lambda}
\def\L{\Lambda}
\def\m{\mu}
\def\n{\nu}
\def\o{\over}
\def\om{\omega}
\def\O{\Omega}
\def\p{\phi}
\def\P{\Phi}
\def\pa{\partial}
\def\pr{\prime}
\def\ra{\rightarrow}
\def\s{\sigma}
\def\S{\Sigma}
\def\t{\tau}
\def\th{\theta}
\def\Th{\Theta}
\def\ti{\tilde}
\def\wti{\widetilde}
\def\jc{J^C}
\def\bj{{\bar J}}
\def\sj{{\jmath}}
\def\bsj{{\bar \jmath}}
\def\bp{{\bar \p}}
\def\faa{Fa\'a di Bruno~}
\def\ca{{\cal A}}
\def\cb{{\cal B}}
\def\cC{{\cal C}}
\def\ce{{\cal E}}
\def\cM{{\cal M}}
\newcommand\sumi[1]{\sum_{#1}^{\infty}}   
\newcommand\fourmat[4]{\left(\begin{array}{cc}  
{#1} & {#2} \\ {#3} & {#4} \end{array} \right)}

%
\def\lie{{\cal G}}
\def\dlie{{\cal G}^{\ast}}
\def\elie{{\widetilde \lie}}
\def\edlie{{\elie}^{\ast}}
\def\hlie{{\cal H}}
\def\wlie{{\widetilde \lie}}
\def\f#1#2#3 {f^{#1#2}_{#3}}
\def\winf{{\sf w_\infty}}
\def\win1{{\sf w_{1+\infty}}}
\def\hwinf{{\sf {\hat w}_{\infty}}}
\def\Winf{{\sf W_\infty}}
\def\Win1{{\sf W_{1+\infty}}}
\def\hWinf{{\sf {\hat W}_{\infty}}}
\def\Rm#1#2{r(\vec{#1},\vec{#2})}          
\def\OR#1{{\cal O}(R_{#1})}           
\def\ORti{{\cal O}({\widetilde R})}           
\def\AdR#1{Ad_{R_{#1}}}              
\def\dAdR#1{Ad_{R_{#1}^{\ast}}}      
\def\adR#1{ad_{R_{#1}^{\ast}}}       
\def\KP{${\rm \, KP\,}$}                 
\def\KPl{${\rm \,KP}_{\ell}\,$}         
\def\KPo{${\rm \,KP}_{\ell = 0}\,$}         
\def\mKPa{${\rm \,KP}_{\ell = 1}\,$}    
\def\mKPb{${\rm \,KP}_{\ell = 2}\,$}    
%
\def\AM#1{A^{(M)}_{#1}}
\def\BM#1{B^{(M)}_{#1}}
\def\Xb{X(b_{M})}
\def\Yb{Y(b_{M})}
\def\Xbo{X_{(0)}(b_{M})}
\def\Ybo{Y_{(0)}(b_{M})}
%
\def\rlx{\relax\leavevmode}
\def\inbar{\vrule height1.5ex width.4pt depth0pt}
\def\IZ{\rlx\hbox{\sf Z\kern-.4em Z}}
\def\IR{\rlx\hbox{\rm I\kern-.18em R}}
\def\IC{\rlx\hbox{\,$\inbar\kern-.3em{\rm C}$}}
\def\one{\hbox{{1}\kern-.25em\hbox{l}}}
\def\0#1{\relax\ifmmode\mathaccent"7017{#1}%
        \else\accent23#1\relax\fi}
\def\omz{\0 \omega}
%
\def\ltimes{\mathrel{\vrule height1ex}\joinrel\mathrel\times}
\def\rtimes{\mathrel\times\joinrel\mathrel{\vrule height1ex}}
%
\def\mark{\noindent{\bf Remark.}\quad}
\def\prop{\noindent{\bf Proposition.}\quad}
\def\theor{\noindent{\bf Theorem.}\quad}
\def\name{\noindent{\bf Definition.}\quad}
\def\exam{\noindent{\bf Example.}\quad}
\def\proof{\noindent{\bf Proof.}\quad}
%
%
\def\PRL#1#2#3{{\sl Phys. Rev. Lett.} {\bf#1} (#2) #3}
\def\NPB#1#2#3{{\sl Nucl. Phys.} {\bf B#1} (#2) #3}
\def\NPBFS#1#2#3#4{{\sl Nucl. Phys.} {\bf B#2} [FS#1] (#3) #4}
\def\CMP#1#2#3{{\sl Commun. Math. Phys.} {\bf #1} (#2) #3}
\def\PRD#1#2#3{{\sl Phys. Rev.} {\bf D#1} (#2) #3}
\def\PLA#1#2#3{{\sl Phys. Lett.} {\bf #1A} (#2) #3}
\def\PLB#1#2#3{{\sl Phys. Lett.} {\bf #1B} (#2) #3}
\def\JMP#1#2#3{{\sl J. Math. Phys.} {\bf #1} (#2) #3}
\def\PTP#1#2#3{{\sl Prog. Theor. Phys.} {\bf #1} (#2) #3}
\def\SPTP#1#2#3{{\sl Suppl. Prog. Theor. Phys.} {\bf #1} (#2) #3}
\def\AoP#1#2#3{{\sl Ann. of Phys.} {\bf #1} (#2) #3}
\def\PNAS#1#2#3{{\sl Proc. Natl. Acad. Sci. USA} {\bf #1} (#2) #3}
\def\RMP#1#2#3{{\sl Rev. Mod. Phys.} {\bf #1} (#2) #3}
\def\PR#1#2#3{{\sl Phys. Reports} {\bf #1} (#2) #3}
\def\AoM#1#2#3{{\sl Ann. of Math.} {\bf #1} (#2) #3}
\def\UMN#1#2#3{{\sl Usp. Mat. Nauk} {\bf #1} (#2) #3}
\def\FAP#1#2#3{{\sl Funkt. Anal. Prilozheniya} {\bf #1} (#2) #3}
\def\FAaIA#1#2#3{{\sl Functional Analysis and Its Application} {\bf #1} (#2)
#3}
\def\BAMS#1#2#3{{\sl Bull. Am. Math. Soc.} {\bf #1} (#2) #3}
\def\TAMS#1#2#3{{\sl Trans. Am. Math. Soc.} {\bf #1} (#2) #3}
\def\InvM#1#2#3{{\sl Invent. Math.} {\bf #1} (#2) #3}
\def\LMP#1#2#3{{\sl Letters in Math. Phys.} {\bf #1} (#2) #3}
\def\IJMPA#1#2#3{{\sl Int. J. Mod. Phys.} {\bf A#1} (#2) #3}
\def\AdM#1#2#3{{\sl Advances in Math.} {\bf #1} (#2) #3}
\def\RMaP#1#2#3{{\sl Reports on Math. Phys.} {\bf #1} (#2) #3}
\def\IJM#1#2#3{{\sl Ill. J. Math.} {\bf #1} (#2) #3}
\def\APP#1#2#3{{\sl Acta Phys. Polon.} {\bf #1} (#2) #3}
\def\TMP#1#2#3{{\sl Theor. Mat. Phys.} {\bf #1} (#2) #3}
\def\JPA#1#2#3{{\sl J. Physics} {\bf A#1} (#2) #3}
\def\JSM#1#2#3{{\sl J. Soviet Math.} {\bf #1} (#2) #3}
\def\MPLA#1#2#3{{\sl Mod. Phys. Lett.} {\bf A#1} (#2) #3}
\def\JETP#1#2#3{{\sl Sov. Phys. JETP} {\bf #1} (#2) #3}
\def\JETPL#1#2#3{{\sl  Sov. Phys. JETP Lett.} {\bf #1} (#2) #3}
\def\PHSA#1#2#3{{\sl Physica} {\bf A#1} (#2) #3}
\def\PHSD#1#2#3{{\sl Physica} {\bf D#1} (#2) #3}
\begin{titlepage}
\vspace*{-1cm}
\noindent
\phantom{bla}  \hfill{\sl BGU-93 / 2 / June - PH} \\
\phantom{bla}
\hfill{hep-th/9306035}
\\
\vskip .3in
\begin{center}
{\large\bf Construction of KP Hierarchies \\
in Terms of Finite Number of Fields \\
and Their Abelianization}
\end{center}
\vskip .3in
\begin{center}
{ H. Aratyn\footnotemark
\footnotetext{Work supported in part by the U.S. Department of Energy
under contract DE-FG02-84ER40173}}
\par \vskip .1in \noindent
Department of Physics \\
University of Illinois at Chicago\\
845 W. Taylor St.\\
Chicago, IL 60607-7059, {\em e-mail}:
u23325@uicvm \\
\par \vskip .3in
{ E. Nissimov$^{\,2}$  and S. Pacheva \footnotemark
\footnotetext{On leave from: Institute of Nuclear Research and Nuclear
Energy, Boul. Tsarigradsko Chausee 72, BG-1784 $\;$Sofia,
Bulgaria. }}
\par \vskip .1in \noindent
Department of Physics, Ben-Gurion University of the Negev \\
Box 653, IL-84105 $\;$Beer Sheva, Israel \\
{\em e-mail}: emil@bguvms, svetlana@bguvms
\par \vskip .3in
\end{center}

\begin{abstract}
The $2M$-boson representations of KP hierarchy are constructed
in terms of $M$ mutually independent two-boson KP representations
for arbitrary number $M$.
Our construction establishes the multi-boson representations of KP
hierarchy
as consistent Poisson reductions of standard KP hierarchy within the
$R$-matrix scheme.
As a byproduct we obtain a complete description of any
finitely-many-field
formulation of KP hierarchy in terms of Darboux coordinates with respect
to the first Hamiltonian structure.
This results in a series of representations of $\Win1\,$
algebra made out of arbitrary even number of boson fields.
\end{abstract}

\end{titlepage}

\noindent
{\large {\bf 1. Introduction}}
\lskip

It has been recognized in the last few years that the integrability
structure
appearing in
the double scaling limit of the one-matrix model can be analyzed in
terms of the KdV
hierarchy augmented by the string equation \ct{Douglas}.
This result created a lot of interest in various types of integrable
hierarchies in connection with attempts to uncover similar pattern in the
multi-matrix models.
However, taking the continuum limit in the multi-matrix
models encountered severe difficulties.
An attempt to circumvent these problems was made in \ct{BX9212}, where
the matrix models were represented  as discrete linear systems giving
rise to lattice
integrable hierarchies from which differential hierarchies were
extracted without taking the continuum limit.
In this approach the one-matrix model resulted in a differential
hierarchy known as
two-boson KP hierarchy \ct{BAK85,2boson}, which via Dirac  constraint
mechanism reduces to simple KdV hierarchy.
In the case of multi-matrix models the same procedure \ct{BX9305} resulted
in pseudo-differential operators, which formally generalized the Lax
operator of two-boson KP hierarchy.
In view of the above development it is, therefore, natural to inquire about
the precise status of these differential operators within the setting
of KP hierarchy, especially, to prove the
Hamiltonian nature of the corresponding flows.

In \ct{ANPV} we addressed the question of linking and classifying the
integrable
systems falling into the general class of \KPl (with $\ell =0,1,2$)
hierarchies
originating in the Adler-Kostant-Symes (AKS) construction \ct{AKS}.
As we show in this paper these considerations will prove to be essential
for the aforementioned matrix models construction.

One of the features of our $R$-matrix  coadjoint-orbit approach was
that it singled out
the two- and four-boson systems as two finite-dimensional \foot{The
terms ``finite-dimensional'' and ``infinite-dimensional'' refer to
number of functional (field) dimensions.}
field representations of KP hierarchy.
As we point out in this paper, the four-boson system has a dual status :
on the one hand -- a finite-dimensional coadjoint orbit
inside the \mKPb hierarchy, and on the other hand -- a composite system
consisting of two independent two-boson systems.
This rises the question whether this picture could be extended,
namely, whether two-boson systems could be used as building blocks
of finitely-many-boson KP hierarchies fitting into the AKS formalism
with Kirillov-Kostant $R$-Poisson bracket. These systems would then
provide finitely-many-field representations of $\Win1$ algebras.
We present in this paper an explicit construction of such systems
consisting of arbitrary finite even number of bosons.
These systems are shown to be legitimate Poisson restrictions of the KP
hierarchy. A crucial role in our approach is played by a
recurrence relation connecting $2M$-boson and $2(M-1)$-boson KP systems.
One can interpret our results as abelianization, meaning that
two-boson KP hierarchies
provide the Darboux coordinates for the many-boson representations of
KP. Each $2M$-boson representation of KP within the first Hamiltonian
structure is built-up out of $M$ mutually commuting two-boson systems.

In section 2 we present the AKS formulation of three integrable
systems
of KP type and their equivalence via symplectic gauge transformations
in a form, which yields the basis of our subsequent construction. More
precisely, in the AKS setting there exist two consistent restrictions of
the KP hierarchy in terms of two- and four-boson systems.
Each of these two systems provides an example of a finitely-many-field
representation of $\Win1$
algebra as it follows automatically by virtue of the symplectic character
of the gauge transformations mapping these systems into the
standard \KPo hierarchy \ct{ANPV}.

The main result of this paper is presented in Section 3, where we prove
that the new class of the multi-boson Lax operators constitutes
a consistent Poisson reduction of
the standard KP manifold with infinitely many fields.
In particular, this results in a series of representations of $\Win1\,$
algebra made out of arbitrary even number of boson fields.
Also, the explicit abelianization formulas for the multi-boson Lax
operators are written down.
Our general construction is illustrated for the specific
case of 6-boson KP hierarchy in Section 4.
We conclude by indicating in Section 5 possible directions of
future investigations.

\lskip
{\large {\bf 2. Algebraic and Geometric Preliminaries}}
\lskip
{\bf 2.1  AKS Approach to KP Hierarchy. Symplectic Gauge
Transformations}
\lskip
We recall first how the AKS \ct{AKS} formalism associates three
KP-type integrable systems labeled by the index $\ell =0,1,2$ to three
possible decompositions of the
Lie algebra $\lie$ of pseudo-differential operators on the circle
into a linear sum of two subalgebras.
Writing an arbitrary pseudo-differential operator $X \in \lie$ as
$X = \sum_{k \geq -\infty} D^k \,X_k (x) \,$ \foot{Throughout the text
$D$ denotes the differential operator $D=\partder{}{x}$, whereas
derivative acting on a function will be denoted by $\, \pa_x f$ .}
we can decompose $\lie$ as
  $\lie = \lie_{+}^{\ell} \oplus
\lie_{-}^{\ell}$ \ct{GKR88,BAK85,ANPV} with:
\be
\lie^{\ell}_{+} = \{\, X_{\geq \ell}
= \sumi{i=\ell} D^i X_i (x) \,\}
\quad;\quad
\lie^{\ell}_{-} = \{\, X_{< \ell}
= \sumi{i=-\ell+1 } D^{-i} X_{-i}(x)\,\}
\lab{subalg}
\ee
 for  $\ell = 0,1,2$.
The corresponding dual spaces with respect to the Adler bilinear
pairing $\, \llangle L \bv X \rrangle = \Tr \( L\, X\) = \int dx \,
{\rm Res} \( L\, X\)\,$ are given by:
\be
{\lie^{\ell}_{+}}^{\ast} = \{ L_{< -\ell}
= \sumi{i=\ell+1} u_{-i}(x) D^{-i}\, \} \;\; ;\;\;
{\lie^{\ell}_{-}}^{\ast} = \{ L_{\geq -\ell} =
\sumi{i=-\ell} u_{i}(x) D^{i}\}
\lab{dsubalg}
\ee
Note the opposite ordering of $D$'s and coefficient functions in
\rf{subalg} and \rf{dsubalg}.
Denoting the projections on the subalgebras in \rf{subalg}
by ${\cal P}_{\pm}^{\ell}$ we can define the $R$-matrix
operator on $\lie$ as
$R_{\ell} \equiv {\cal P}_{+}^{\ell}  - {\cal P}_{-}^{\ell} $.
There exists a new Lie commutator on $\lie$ associated
to each $R_{\ell}$-matrix and
defined by $\lb X , Y \rb_{R_{\ell}}
\equiv \lb R_{\ell} X , Y \rb/2 + \lb X, R_{\ell} Y \rb /2
=\lb X_{\geq \ell},
Y_{\geq \ell}\rb - \lb X_{< \ell} , Y_{< \ell} \rb$.
The Poisson structure on $\dlie$ follows now naturally by generalizing
the Kirillov-Kostant formula to the $R_{\ell}$-commutator as follows:
\be
\{ F \, , \, H \}_{R_{\ell}} (L) =
- \llangle L \bv {\sbr{\nabla F (L)}{\nabla H (L)}}_{R_{\ell}}
\rrangle   \lab{Rbra}
\ee
see \ct{AKS,ANPV} for details. The $\,R_{\ell}$-Poisson bracket
\rf{Rbra} is the first Hamiltonian structure for \KPl
hierarchy.

Consider now Casimir functions on $\dlie$ defined as functions, which
are invariant
under coadjoint action of the corresponding Lie group $G$.
The Casimir functions constitute a set of functions in involution on
the Poisson manifold .
A convenient choice of Casimirs is provided by
$H_{n+1} = { 1 \o n+1} \Tr L^{n+1}$ for which $\nabla H_{n+1}=
 (L^n)_{\ge \ell}$.
The Hamiltonian equations of motion on
$\( \dlie, \{ \cdot ,\cdot\}_{R_{\ell}} \)$
associated to these Casimir functions :
\be
\partder{L}{t_n} = {\pbr{H_{n }}{L}}_{R_{\ell}}
\lab{hamflow}
\ee
take, according to \rf{Rbra}, the form of Lax
evolution equations  on $ \dlie$
for all three integrable \KPl systems:
\be
\partder{L}{t_n} = \sbr{(L^n)_{\ge \ell}}{L} \qquad \ell
= 0, 1, 2       \lab{aksflow}
\ee

There is a way of relating Lax operators of different \KPl ~hierarchies
by a map,
which plays a role of gauge transformation.
Consider first \rf{aksflow} with $\ell=0$.
It describes the standard KP flow equation with the Lax operator:
\be
L \equiv D +  \sumi{i=1} u_i (x, t) D^{-i}
\lab{laxop}
\ee
with the first Hamiltonian structure induced by the $R_0$-bracket
\rf{Rbra} being
$\{ u_n(x)\, , \, u_m(y) \}_{R_0} = \Omega^{(0)}_{n-1,m-1}(u(x)) \, \d
(x-y)$ ,
where the Watanabe form on the right hand side can be obtained from
the general expression:
\be
\Omega^{(\ell)}_{nm}(u(x)) \equiv \sum_{k=0}^{n+\ell}(-1)^k
{n+ \ell\choose k} u_{n+m+\ell-k+1}(x) D^k_x -
\sum_{k=0}^{m+\ell} {m+\ell\choose k} D^k_x
u_{n+m+\ell-k +1}(x)     \lab{omega}
\ee

We call the integrable system characterized by $\ell=1$ in \rf{aksflow}
a \mKPa hierarchy and associate to it a Lax operator as follows.
Consider elements in
${{\cal G}_{-}^{\ell=1}}^{\ast} $ \rf{dsubalg} of the type
${\wti L}_{1} = D + u_0 + \, u_1 D^{-1} $, which
preserve their form under $R_1$-coadjoint action,
spanning therefore a $R_1$-orbit of finite field dimensions.
A complete Lax operator is obtained by adding ${\wti L}_{1}$ to the
general element $L_{-}$ of ${{\cal G}_{+}^{\ell=1}}^{\ast} $
\rf{dsubalg} :
\be
L^{(\ell=1)}= {\wti L}_{1}+ L_{-} = D + u_0 +
\, u_1 D^{-1} + \sum_{i \geq 2} u_{i}\,D^{-i}.
\lab{mkpalax}
\ee
There is a map, resembling a gauge transformation, between the Lax
operators $L$ \rf{laxop} and $L^{(\ell=1)}$  \rf{mkpalax} :
\be
L \equiv G^{-1} L^{(\ell=1)} G  = D +  \sumi{i=1} v_i D^{-i} \qquad;
\qquad
G \equiv \exp \( - \int^x u_0\, dx^{\pr} \)
\lab{gaug1}
\ee
Finally, we consider the \mKPb hierarchy. Here elements of ${\cal
G}_{-}^{{\ell =2}\,\ast}\, $ \rf{dsubalg} of the form
\be
{\wti L}_{2} = u_{-1} D + u_0 + u_1 D^{-1} + u_2 D^{-2}  \lab{L2}
\ee
span an invariant subspace under the coadjoint action
induced by $R_{\ell=2}$-matrix.
The complete Lax operator for \mKPb is then given by
$L^{(\ell=2)} = {\wti L}_{2} + L_{-} =
u_{-1} D + u_0 + u_1 D^{-1} + u_2 D^{-2} +
\sum_{i \ge 3} u_i D^{-i} $ and transforms to the Lax operator of
\mKPa hierarchy
under the gauge transformation generated by the centerless Virasoro
group.
Explicitly we find \ct{ANPV}:
\be
e^{\p (x) D} L^{(\ell=2)} e^{- \p (x) D} =  D + {\ti u}_0 +
{\ti u}_1 D^{-1} + {\ti u}_2 D^{-2} + \sum_{i \ge 3} {\ti u}_i D^{-i}
\lab{gaug2}
\ee
where $ \p (x)$ is chosen in such a way that $u_{-1} \( F_{\p} (x) \)
= \pa_x F_{\p} (x) $ with $ F_{\p} (x) = \exp ( \p (x) \pa_x) x $
representing a finite conformal transformation.
Clearly, the Lax operator on the right hand side of \rf{gaug2} belongs
to \mKPa hierarchy.

The main result of \ct{ANPV} was an explicit proof that the gauge
transformations
in \rf{gaug1} and
\rf{gaug2} are symplectic maps, meaning that they map the
$R_{\ell}$-Poisson bracket structure for \KPl to the
$R_{{\ell}^{\pr}}$-bracket structure for ${\rm\,KP}_{\ell^{\pr}}\,$.
This result established full gauge equivalence between all three
integrable systems described by the \KPl hierarchies.
We will illustrate this principle for the finite-dimensional
cases of two- and four-boson systems associated with
${\wti L}_1$ and ${\wti L}_2$ \rf{L2} operators.
\lskip
{\bf 2.2 Two- and Four-Boson Representations of KP and $\bf \,\Win1$
and Their Relation}
\lskip
The starting point is the two-boson Lax operator ${\wti L}_{1} =
D + b + \, a D^{-1} $ in the \mKPa hierarchy (notations for Lax
coefficients are changed
for later convenience). The corresponding $R_1$-bracket reads
$ \,\{ \, a  (x)\, , \, b (y) \}_{R_1} = -  \pa_x \d (x -y) \,$
and zero otherwise.

Under the gauge transformation:
\be
L_1 \equiv e^{\int b } {\wti L}_1 e^{-\int b } = D + a \( D-b\)^{-1} =
D +  \sumi{n=0} (-1)^n a P_n (-b) D^{-1-n}
\lab{gaugtwo}
\ee
${\wti L}_1$ transforms to the constrained \KPo
Lax operator $L_1$ with coefficients $u_{n+1} =
(-1)^n a P_n (-b) $ given in terms of the \faa polynomials
$P_n (b) \equiv (\pa + b )^n \cdot 1$.
As a consequence of the symplectic character of the gauge
transformation in \rf{gaugtwo} we find therefore
\be
\Bigl\{ (-1)^n a P_n (-b) \, , \, (-1)^m a P_m(-b) \Bigr\}_{R_0} =
\Omega^{(0)}_{nm}\Bigl( u_{n+1} = (-1)^n a P_n (-b) \Bigr) \, \d (x-y)
\lab{twowatanabe}
\ee
and hence the two-boson system realizes the $\,\Win1$ algebra.

Similar remarks apply to the restricted KP system of four
bosons
described by the Lax operator ${\wti L}_2$ \rf{L2} of the \mKPb hierarchy.
The relevant gauge transformation \rf{gaug2} acts now as follows
($F^{\pr}_{\p} \equiv \pa_x F_{\p}$) :
\br
e^{\p (x) D} {\wti L}_2  e^{- \p (x) D} \eq  D +  u_{0} \(F_{\p} (x) \)+
 u_1 \(F_{\p} (x) \) D^{-1} F^{\pr}_{\p} (x)  +
 u_2 \(F_{\p} (x) \) D^{-1} F^{\pr}_{\p} (x) D^{-1} F^{\pr}_{\p}(x)
 \nonu   \\
\eq e^{ -\ln F^{\pr}_{\p} (x)} \( D + B_2 + A_2 D^{-1} + A_1 (D-{\bar
B}_1)^{-1}
D^{-1}\) e^{\ln F^{\pr}_{\p} (x)} \lab{gaugfour}
\er
Therefore, it connects  ${\wti L}_2$ via additional Abelian gauge
transformation to the \mKPa Lax operator
\be
{\hat L}_2 = D + B_2 + A_2 D^{-1} + A_1 (D-{\bar B}_1)^{-1} D^{-1}
\lab{l2kpone}
\ee
whose coefficient fields
\br
A_1 \eq u_2 \(F_{\p} (x) \) \(F^{\pr}_{\p} (x) \)^2 \quad,\quad
{\bar B}_1 = - \pa_x \ln F^{\pr}_{\p} (x) \lab{abtwo} \\
A_2 \eq u_1 \(F_{\p} (x) \) F^{\pr}_{\p} (x)
\quad,\quad B_2 = u_0 \(F_{\p} (x) \) - \pa_x \ln F^{\pr}_{\p}
(x)
\nonu
\er
satisfy according to \ct{ANPV} the following algebra :
\br
\lcurl A_2 (x) \,, \,B_2 (y) \rcurl \eq - \pa_x \d (x-y)
\lab{4pb-a}   \\
\lcurl A_1 (x)\, , \,{\bar B}_1 (y) \rcurl \eq - \( \pa_x + {\bar B}_1
(x)\)\, \pa_x  \d (x-y)  \lab{4pb-b}  \\
\lcurl A_1  (x) \,, \, A_1 (y) \rcurl \eq
- 2 A_1  (x) \pa_x \d (x-y)  - \( \pa_x A_1\) \d (x-y) \lab{4pb-c}
\er
as follows from the original $R_2$-Poisson brackets \rf{Rbra} for
$u_{-1}, u_0, u_1, u_2$ within the \mKPb hierarchy.

In order to end up with Lax operator in \KPo we then apply
to \rf{l2kpone} gauge transformation generated by $-\int B_2$ with
the result:
\br
L_2 \eq e^{ \int B_2 }  {\hat L}_2 e^{ -\int B_2 } =
D  + A_2 \(D- B_2\)^{-1} + A_1 (D-B_1)^{-1} \(D- B_2\)^{-1}
\lab{abab}\\
\eq  D +  \sumi{n=0} (-1)^n A_2 P_n (-B_2) D^{-1-n} +
\sumi{n=0} A_1 P^{(2)}_n (B_2 ,B_1 ) D^{-2-n}
\equiv D +  \sumi{k=1} U_k \lb A_{1,2},B_{1,2}\rb  D^{-k} \nonu
\er
where $B_1 \equiv {\bar B}_1 +B_2$ and where we have introduced
the double \faa polynomials
$P^{(2)}_n (B_2 ,B_1 )= \sum_{l,k \geq 0}^{l+k=n} ( -\pa + B_1)^l
(-\pa + B_2)^k \cdot 1$ (cf. eqs.(44)-(51) in ref.\ct{ANPV}).
On basis of a theorem \ct{ANPV} about the symplectic character of
both types of gauge
transformations used in \rf{gaugfour} and \rf{abab} we know that
coefficient fields
of $L_2$ from \rf{abab} satisfy the Poisson algebra :
\be
\Bigl\{ U_n \lb A_{1,2},B_{1,2}\rb (x)\, , \,
U_m \lb A_{1,2},B_{1,2}\rb (y) \Bigr\}_{R_0} =
\Omega^{(0)}_{n-1,m-1}\Bigl( U_k \lb A_{1,2},B_{1,2}\rb \Bigr) \,
\d (x-y)  \lab{W4}
\ee
whenever $\, A_{1,2},\, B_{1,2}$ satisfy \rf{4pb-a}-\rf{4pb-c}
and, therefore, the four-boson system
forms a representation of $\,\Win1$ algebra.

As already observed in \ct{ANPV} the four-boson Poisson algebra
\rf{4pb-a}-\rf{4pb-c} decomposes into direct sum of Heisenberg algebra
generated by the
two-boson system $(A_2,B_2)$ and separate algebra of coupled spin-2 and
spin-1 fields $(A_1,{\bar B}_1)$.
It is well-known (see for instance \ct{miura}) that there exists in the
KP setting a
generalized Miura transformation, which maps elements of the Heisenberg
algebra
to the higher spin algebras.
In the case of $(A_1,{\bar B}_1)$ fields and their algebra the generalized
Miura
transformation
takes the following form:
\be
A_1 = \( \pa + b_1 \) a_1  \qquad,\qquad {\bar B}_1 = b_1
\lab{gmfour}
\ee
in terms of two-boson system $(a_1,b_1)$ satisfying the Heisenberg
algebra
$\{ a_{1}(x)\,, \,b_{1}(y)\}= - \pa_x \d (x-y)$.
Summarizing we can say that the four-boson KP system given by
\rf{abab} can be abelianized in terms of two mutually
Poisson-commuting two-boson systems $(a_1,b_1)$ and $(a_2,b_2)$
entering into the generalized Miura transformation:
\br
A_2 = a_2  \quad,\quad A_1 = \( \pa + b_1 \) a_1 \quad &,& \quad
B_2 = b_2 \quad,\quad  B_1 = b_1 +b_2         \lab{ggmfour} \\
\lcurl a_i (x)\, ,\, b_j (y) \rcurl \eq - \d_{ij} \pa_x \d (x-y)
\qquad  i,j=1,2 \lab{H2}
\er
which reproduces the algebra \rf{4pb-a}-\rf{4pb-c}.
The application of the generalized Miura transformation \rf{ggmfour}
can be visualized as a recurrence
relation connecting the two-boson $L_1 \equiv D + a_1 (D-b_1)^{-1}$
and four-boson $L_2\,$ \rf{abab} Lax operators. Using \rf{ggmfour},
we find by simple calculation :
\be
L_2 = e^{\int b_2} \Bigl\lb b_2 + (a_2 -a_1 )D^{-1} +
D L_1 D^{-1} \Bigr\rb e^{-\int b_2}
\lab{dres1}
\ee
This recurrence relation connects two- and four-boson Lax operators by
an Abelian gauge transformation and the dressing operation $D L D^{-1}$.
This will remain a general feature in the Section 3 when we address
the problem of building
multi-boson KP hierarchies out of any number of independent two-boson
systems.
We will use there this relation to study the Poisson bracket algebra
of the composite systems.

\lskip
{\bf 2.3 Poisson Reduction}
\lskip
The 4-boson representation of \KP hierarchy admits an alternative
description
in terms of Poisson reduction on the phase space of general Lax operators
\rf{laxop}. It is precisely the Poisson reduction scheme which
provides the proper basis for our
generalization of the previous construction of 2-boson and 4-boson \KP
representations to representations of \KP in terms of arbitrary finite
number of boson field pairs.

First, let us recall some general notions \ct{MR86}. Let
$\, \( \cM ,P\) \,$ be a smooth Poisson  \foot{$\cM\,$ needs not
to be symplectic manifold, i.e., the Poisson structure $P\,$ might be
degenerate (the Poisson tensor $\om^{ij}(x)\,$ being non-invertible).}
manifold with Poisson structure $\, P \; : \; T^{\ast}(\cM ) \longrightarrow
T(\cM )$ . In local coordinates $\, \lcurl x^i \rcurl_{i=1}^{{\rm dim}\cM}
\,$ on $\cM\,$ the Poisson bracket of arbitrary smooth functions,
defined by the Poisson structure $P$ , is given as :
\be
\lcurl f ,\, g \rcurl_P = \llangle P \grad f \bv \grad g \rrangle
= \om^{ij}(x) \partder{f}{x^i} \partder{g}{x^j}   \lab{p1}
\ee
where the angle brackets denote pairing between $\,T^{\ast}(\cM )\,$ and
$\, T(\cM )\,$.

Let $S$ be a smooth submanifold of $\cM\,$ with local coordinates
$\, \lcurl \s^\a \rcurl_{\a =1}^{{\rm dim} S} $ and embedding
$\, \m \; :\; S \longrightarrow \cM $ . Now, a Poisson structure
$\, P^{\pr}\; : \; T^{\ast}(S) \longrightarrow T(S)\,$ on
$\, S \subset \cM\,$ is called {\em Poisson reduction} of
$\, P\,$ if for arbitrary
functions on $\cM\,$ the following property is satisfied :
\be
\m^{\ast} \( \lcurl f ,\, g \rcurl_P \) =
\lcurl \m^{\ast}f ,\, \m^{\ast}g \rcurl_{P^{\pr}}  \lab{p2}
\ee
In other words, restriction of the Poisson brackets w.r.t. $P$ of
arbitrary functions on $\cM\,$ to the submanifold $S\,$ is equivalent to
computing the Poisson brackets w.r.t. $P^{\pr}$ of the restrictions on
$S\,$ of these same functions \foot{Let us stress that, in the case when
$\, \( \cM ,P\) \,$ is symplectic, the {\em Poisson} reduction
$P^{\pr}$ of the Poisson structure $P\,$ is in general different from
the {\em Dirac} reduction thereof. The associated Dirac brackets
are of the form :
$ \, \lcurl \m^{\ast}f ,\, \m^{\ast}g \rcurl_{DB} =
\m^{\ast} \Bigl( \lcurl f ,\, g \rcurl_P  -
\lcurl f, \Psi_A \rcurl_P \(\cC^{-1}\)^{AB} \lcurl \Psi_B , g \rcurl_P
\Bigr)\,$
where $ S = \lcurl \Psi_A =0 \rcurl \,$ is defined through the set of
Dirac second class constraints $\Psi_A\,$ and $\, \cC^{AB} =
\lcurl \Psi_A ,\, \Psi_B \rcurl_P $.}.

In local coordinates eq.\rf{p2} can be written as (recall
$\m^{\ast}f (\s ) = f \bigl( x(\s ) \bigr)$ ) :
\be
\om^{ij}\bigl(x(\s )\bigr) =
{\hat \om}^{\a\b}(\s ) \partder{x^i}{\s^\a} \partder{x^j}{\s^\b}
\;\; \Bigl( \;\; = \lcurl x^i (\s ), x^j (\s ) \rcurl_{P^{\pr}}
\Bigr)   \lab{p3}
\ee
where $ {\hat \om}^{\a\b}(\s ) $ is the Poisson tensor of $P^{\pr}$ .

Comparing \rf{p3} with \rf{W4} and identifying $\, x^i \sim
L$  from \rf{laxop}, $ \s^\a \sim \( A_{1,2}, B_{1,2} \) $ and
$\, x^i (\s ) \sim L_2 \,$ from \rf{abab}, it is readily
seen that 4-boson representation of \KP \rf{abab},\rf{W4}
is indeed a genuine Poisson
reduction of the original Kirillov-Kostant $R_0$-Poisson structure
$\, P\,$ \rf{Rbra} on the infinite-dimensional Lax manifold
$ \,\cM = \lcurl L = D + \sum_{k=1}^{\infty}
u_k (x) D^{-k} \rcurl\,$ to the Poisson structure $\, P^{\pr}\,$
\rf{4pb-a}-\rf{4pb-c} on the finite-dimensional manifold
$ S = \Bigl\{ L_2 \( A_{1,2}, B_{1,2} \) \Bigr\} \,$
\rf{abab}. Obviously, similar remark applies as well to the 2-boson KP
system.

\lskip
{\large {\bf 3. $\bf KP$ and $\bf W_{1+\infty}$ in Terms of $\bf 2M$
Fields for Arbitrary $\bf \, M$}}
\lskip
Let us consider the sequence of pseudo-differential operators
obtained from the natural generalization of the recursive
relation eq.\rf{dres1} for arbitrary $M=2,3,\ldots$
\br
L_{M} &\equiv& L_M (a,b) \equiv L_M \(a_1 ,b_1 ; \ldots
; a_M ,b_M \)  \nonu  \\
L_{M} \eq e^{\int b_{M}} \Bigl\lb b_{M} +
(a_{M} - a_{M-1} )D^{-1} + D L_{M-1} D^{-1} \Bigr\rb
e^{-\int b_{M}}                   \lab{3-1}
\er
where $L_1$ and $L_2$ are the 2- and 4-boson \KP operators,
respectively, (eqs.\rf{gaugtwo} and \rf{abab}, or \rf{dres1}),
and the boson fields $\, \( a_r , b_r \)_{r=1}^M \,$ span
Heisenberg Poisson bracket algebra :
\be
\lcurl a_r (x),\, b_s (y) \rcurl_{P^{\pr}} =
 - \d_{rs} \pa_x \d (x-y)          \lab{3-2}
\ee
Using the identities $\, e^{\int b_{M}} D^{\pm 1} e^{-\int b_{M}} =
\( D - b_{M}\)^{\pm 1} $,
eq.\rf{3-1} can be rewritten in an equivalent form, similar to
expression \rf{abab} for $L_2$, which is valid for any $M$ :
\be
L_{M} = D + \sum_{l=1}^{M} \AM{l}
\( D - \BM{l}\)^{-1} \( D - \BM{l+1}\)^{-1} \cdot\cdot\cdot
\( D - \BM{M}\)^{-1}    \lab{3-3}
\ee
where the coefficient fields satisfy the simple recursion
relations :
\br
\AM{M} \eq a_M \quad, \quad \BM{M} = b_M \quad ,\quad
\BM{l} = b_M + B^{(M-1)}_l \quad \; (l=1,2,\ldots ,M-1)
\lab{recur}  \\
A_{1}^{(M)}\eq \( \pa + B_{1}^{(M-1)}\)  A_{1}^{(M-1)} \; ,\quad
\AM{l} = A_{l-1}^{(M-1)} + \( \pa + B_l^{(M-1)} \) A_l^{(M-1)}
\quad  (l=2,\ldots ,M-1)  \nonu
\er
These recursion relations can be easily solved in terms of the
free fields $\, a_r , b_r\,$ from \rf{3-2} to yield :
\br
\BM{l} \eq \sum_{s=l}^{M} b_s \qquad \qquad , \qquad \qquad
\AM{M} = a_{M}
\lab{3-4}  \\
\AM{M-r} \eq \sum_{n_r =r}^{M-1} \cdot\cdot\cdot \sum_{n_2 =2}^{n_3 -1}
\sum_{n_1 =1}^{n_2 -1} \( \pa + b_{n_r} + \cdot\cdot\cdot +
b_{n_r -r +1} \) \cdot\cdot\cdot \( \pa + b_{n_2} + b_{n_2 -1}\)
\( \pa + b_{n_1}\) a_{n_1} \lab{3-5}
\er
The coefficients of the pseudo-differential operator \rf{3-1}
(or \rf{3-3}) have the following explicit expressions :
\br
L_{M}\eq D + \sum_{k=1}^{\infty}
U_k \lb (a,b) \rb (x) D^{-k}  \lab{3-A} \\
U_k \lb (a,b) \rb (x) \eq a_{M} P^{(1)}_{k-1}\( b_{M}\) +
\sum_{r=1}^{\min (M-1,k-1)} \AM{M-r} P^{(r+1)}_{k-1-r}
\bigl( b_{M}, b_{M} + b_{M-1}, \ldots ,
\sum_{l=M-r}^{M} b_l \bigr)   \lab{3-B}
\er
where $\,\AM{M-r}$ are the same as in \rf{3-5}, and
$\, P^{(N)}_n \,$ denote the (multiple) \faa polynomials :
\be
P^{(N)}_n (B_N ,B_{N-1},\ldots ,B_1) =
\sum_{m_1 + \cdot\cdot\cdot + m_N = n}
\( -\pa + B_1 \)^{m_1} \cdot\cdot\cdot \( -\pa + B_N\)^{m_N}
\cdot 1   \lab{multifaa}
\ee

The main result of this section is contained in the following :

 \theor {\em The $2M$-field Lax operators
\rf{3-1} (or \rf{3-3}) are consistent Poisson reductions
of the general KP Lax operator
\rf{laxop} for any $\, M=1,2,3,\ldots \,$} .

In other words, we shall prove that the Heisenberg Poisson bracket algebra
$\, P^{\pr}$ \rf{3-2} for
$\, \( a_r ,b_r \)_{r=1}^{M} $ {\em implies} the following Poisson
brackets for $L_{M}$ from \rf{3-3} or \rf{3-1}
(recall eqs.\rf{p2} and \rf{p3}) :
\be
\Bigl\{ \llangle L_{M} \bv X \rrangle \, ,\,
\llangle L_{M} \bv Y \rrangle \Bigr\}_{P^{\pr}} =
- \llangle L_{M} \bv \left\lb X,\, Y
\right\rb \rrangle    \lab{3-6}
\ee
where $\, X,\, Y\,$ are arbitrary fixed elements of the algebra
of pseudo-differential operators and $\, \langle \cdot \v \cdot
\rangle\,$ indicates the Adler bilinear pairing.

\mark
Let us particularly stress that the truncation of the form of the
general Lax operator \rf{laxop} within the original KP Poisson
brackets \rf{Rbra}, leading to \rf{3-6},
may {\em not} be necessarily consistent. Namely, it
does {\em not} automatically guarantee
the closure of the infinite number of Poisson brackets for
the {\em infinite} number of Lax coefficient fields
$\, U_k \lb (a,b) \rb (x)\,$ \rf{3-B} as
functionals of the {\em finite} number of independent fields
$\, (a_r ,b_r )\,$ w.r.t.
their fundamental Poisson brackets \rf{3-2}.

Thus, the present proof
that eq.\rf{3-6} is a consistent Poisson reduction (cf. subsection
2.3) provides the principle ingredient in the construction of
integrable Hamiltonian systems which are representations
of \KP hierarchies in terms of finite number of fields
\foot{Flow equations with Lax operators of the form \rf{3-3} recently
appeared in the study of multi-matrix models \ct{BX9305}. Our theorem
proves that these flows are Hamiltonian ones.}.

The proof of \rf{3-6} proceeds by induction in $\, M$ . It has
already been established for $M=1$ \ct{BAK85,2boson} and
$ M=2 $ \ct{ANPV}. Now, let
us assume that \rf{3-6} is valid for $\, L_{M-1}$ and rewrite
$\, \llangle L_{M} \bv X \rrangle\,$ in the form :
\br
\llangle L_{M} \bv X \rrangle \eq \int dx \,\( a_{M} - a_{M-1} \)
(x) \Xbo (x) + \llangle L_{M-1} \bv \( D^{-1} \Xb D \)_{+}
\rrangle   \lab{3-7}  \\
\Xb &\equiv& e^{-\int b_{M}} X e^{\int b_{M}} \lab{3-7a}
\er
where the subscripts $(+)$ and $(0)$ denote
purely differential and zero order part of the corresponding
(pseudo-)differential symbol. Substituting \rf{3-7}
into the l.h.s. of \rf{3-6} we get
\br
& & \Bigl\{ \llangle L_M \bv X \rrangle \, ,\,
\llangle L_M\bv Y \rrangle \Bigr\}_{P^{\pr}} = \nonu  \\
& &\int\int dx \, dy \, \lcurl \Xbo (x) \( a_{M} - a_{M-1} \) (x) \, ,\,
\Ybo (y) \( a_{M} - a_{M-1} \) (y) \rcurl_{P^{\pr}}  \nonu \\
&+& \lcurl \llangle L_{M-1} \bv \( D^{-1} \Xb D \)_{+}
\rrangle \, ,\, \llangle L_{M-1} \bv \( D^{-1} \Yb D \)_{+}
\rrangle \rcurl_{P^{\pr}}   \nonu \\
&+& \int dy \( \, \Xbo (y) \left\lb \llangle L_{M-1} \bv
\( D^{-1} \lcurl a_{M}(y)\, ,\, \Yb \rcurl_{R^{\pr}} D\)_{+}
\rrangle  \right.\right.   \nonu  \\
&-&  \left.\left.
\llangle\lcurl a_{M-1} (y)\, ,\, L_{M-1}
\rcurl_{P^{\pr}} \bv \( D^{-1} \Yb D \)_{+} \rrangle \right\rb
- \Bigl( \Xb \longleftrightarrow \Yb \,\Bigr)\,\)
 \lab{3-8}
\er
Using \rf{3-2} and \rf{3-7a} one can easily check the identities :
\br
\lcurl a_{M}(y)\, ,\, \Xb \rcurl_{P^{\pr}} \eq
- \left\lb \d (y-x)\, ,\, \Xb \right\rb   \lab{3-9}  \\
\lcurl a_{M-1} (y)\, ,\,  L_{M-1} \rcurl_{P^{\pr}} \eq
\left\lb \d (y-x)\, ,\, \( L_{M-1}\)_{-} \right\rb       \lab{3-10}
\er
where the r.h.s. of \rf{3-9} and \rf{3-10} indicate
(pseudo-)differential operator commutators w.r.t. $\, x$ , and the
subscript $(-)$ denotes purely pseudo-differential part.

Using the induction hypothesis for the second term in the r.h.s.
of \rf{3-8} and
substituting \rf{3-9} and \rf{3-10} into \rf{3-8} we obtain :
\br
& & \Bigl\{ \llangle L_{M}\bv X \rrangle \, ,\,
\llangle L_{M}\bv Y \rrangle \Bigr\}_{P^{\pr}} =
- \int dx \, \bigl( a_{M} - a_{M-1} \bigr)(x) \left\lb \Xb , \Yb
\right\rb_{(0)}(x)    \nonu   \\
&-&\llangle L_{M-1} {\bigg\v} \left\lb
\( D^{-1} \Xb D \)_{+} , \( D^{-1} \Yb D \)_{+} \right\rb
\right. \nonu  \\
&+&\left.  \biggl( D^{-1} \( \left\lb \Xbo ,\Yb \right\rb +
\left\lb \Xb,\Ybo \right\rb \) D \biggr)_{+} \right.  \nonu  \\
&-& \left.\left\lb \Xbo , \( D^{-1} \Yb D \)_{+} \right\rb -
\left\lb \( D^{-1} \Xb D \)_{+} , \Ybo \right\rb  \rrangle =
\nonu \\
& -& \int dx \, \bigl( a_{M} - a_{M-1} \bigr)(x) \left\lb \Xb ,\Yb
\right\rb_{(0)}(x) - \llangle L_{M-1} \bv \left\lb
D^{-1} \Xb D \, ,\, D^{-1} \Yb D  \right\rb_{+} \rrangle  \nonu \\
&=& - \llangle L_{M} \bv \lb X,Y \rb \rrangle  \lab{3-11}
\er
where in the last equality once again representation \rf{3-7}
was used. This completes the proof of our main statement about
the consistency of the \KP Poisson reduction \rf{3-6}.

Let us point out the following important observation. Eqs.\rf{3-4}
and \rf{3-5} are nothing but {\em abelianization} of the
$\, 2M$-field \KP hierarchy given by \rf{3-3}. Namely, all
coefficients $\, U_k \lb (a,b) \rb (x)$ of the
$\, 2M$-field \KP Lax operator \rf{3-A} are explicitly
expressed (see eq.\rf{3-B}) in terms  of $\,M \,$ pairs of
free fields $\, \( a_r ,b_r \)_{r=1}^{M}$ satisfying the
Heisenberg Poisson bracket algebra \rf{3-2}. From general
Hamiltonian point of view $\, \( a_r ,b_r \)_{r=1}^{M}$ can be
viewed as Darboux canonical coordinates on the phase space of the
$\, 2M$-field \KP system.

Furthermore, eq.\rf{3-6} provides us with explicit (Poisson bracket)
realization of $\Win1$ algebra in terms of $\, 2M$ bosons for
any $\, M=1,2,3,\ldots $ . Indeed, according to \rf{3-6}
the coefficient fields \rf{3-A} satisfy, as functionals of
$\, \( a_r ,b_r \)_{r=1}^{M}$ , the $\Win1$ Poisson bracket algebra
w.r.t. $P^{\pr}$ \rf{3-2} :
\be
\Big\{ U_k \lb (a,b) \rb (x) \, ,\, U_l \lb (a,b) \rb (y) \Bigr\}_{P^{\pr}}
= \O^{(0)}_{k-1,l-1} \bigl( U \lb (a,b) \rb \bigr) \d (x-y) \lab{3-C}
\ee
where $\, \O^{(0)}_{kl}\,$ is given in \rf{omega}.

\lskip
{\large {\bf 4. Example: 6-Boson $\bf KP$ Hierarchy}}
\lskip
In this section we shall specialize the general formulae of the previous
section to represent the  KP hierarchy in term of 6 boson fields.
The Lax operator \rf{3-3} for $\, M=3$  is :
\br
L_{(3)} \eq D + A_3\, (D - B_3 )^{-1} + A_2\, (D - B_2 )^{-1}
(D - B_3)^{-1} \nonu \\
&+& A_1 \,(D - B_1 )^{-1}(D - B_2 )^{-1}(D - B_3 )^{-1}  \lab{4-1}
\er
It is abelianized by the substitutions (cf. \rf{3-4},\rf{3-5}) :
\br
B_1 \eq b_1+b_2+b_3 \;\;\; ,\;\;\; B_2 = b_2+b_3 \;\;\; ,\;\;\;
B_3 = b_3  \lab{4-2} \\
A_1 \eq (\pa + b_1 + b_2 ) (\pa + b_1 )a_1 \;\; ,\;\;
A_2 = (\pa + b_1 )a_1 +  (\pa + b_2 )a_2 \;\; ,\;\; A_3 = a_3
\lab{4-3}
\er
The fields $\, \(A_r ,B_r \)_{r=1}^3$ satisfy the Poisson bracket
algebra (below we use notations $\, {\bar B}_{1,2} = B_{1,2} -
B_3$ ) :
\br
\lcurl A_3 (x) , B_3 (x) \rcurl \eq -\pa_x \d (x-y)  \lab{4-A} \\
\lcurl A_2 (x) , A_2 (y) \rcurl \eq -2 A_2 (x) \pa_x \d (x-y) -
 \(\pa_x A_2 \)\, \d (x-y)  \lab{4-A1} \\
\lcurl A_2 (x) , A_3 (y) \rcurl \eq - 3 A_3 (x) \pa_x \d (x-y) -
 2 \(\pa_x A_3 \)\, \d (x-y)  \lab{4-B} \\
\lcurl A_2 (x) , {\bar B}_2 (y) \rcurl \eq - \(\pa_x  +
 {\bar B}_2 (x)\) \pa_x \d (x-y)      \lab{4-C1}  \\
\lcurl A_2 (x) , {\bar B}_1 (y) \rcurl \eq - \( 2\pa_x  +
 {\bar B}_1 (x) \) \pa_x \d (x-y)       \lab{4-C} \\
\lcurl A_1 (x) , {\bar B}_1 (y) \rcurl \eq - \( \pa_x  +
 {\bar B}_1 \) \( \pa_x  + \({\bar B}_1   - {\bar B}_2 \) \)
 \pa_x \d (x-y)       \lab{4-D} \\
\lcurl A_1 (x) , A_1 (y) \rcurl \eq
A_1 (x) \( \pa_x - {\bar B}_1 \)^2 \d (x-y) -
\( \pa_x + {\bar B}_1 \)^2 A_1 \, \d (x-y)    \nonu  \\
&+& \(\pa_x \( A_1 {\bar B}_2 \)\) \d (x-y)  +
2 A_1 {\bar B}_2 \,\pa_x \d (x-y)     \lab{4-E}
\er
Finally, the $\Win1$ fields in the 6-boson realization
$\, \(A_r ,B_r \)_{r=1}^3$ or, equivalently in terms of the
Darboux fields $\, \(a_r ,b_r \)_{r=1}^3$, read:
\br
U_1 \eq A_3 = a_3 \qquad\qquad\qquad\qquad\qquad\qquad\qquad\qquad
\qquad\qquad ({\rm spin ~1}) \lab{4-4}  \\
U_2 \eq A_3 B_3 + A_2 = a_3 b_3 + (\pa + b_1 )a_1 +
(\pa + b_2 )a_2   \qquad\qquad ({\rm spin ~2})      \lab{4-5} \\
U_s \eq A_3 P_{s-1}^{(1)}(B_3) + A_2 P_{s-2}^{(2)}(B_3, B_2)
+ A_1 P_{s-3}^{(3)} (B_3, B_2, B_1)    \nonu   \\
\eq a_3 P_{s-1}^{(1)}(b_3) +\left\lb (\pa +b_1 )a_1 +
(\pa + b_2 )a_2 \right\rb P_{s-2}^{(2)} (b_3 , b_3 + b_2 ) \nonu  \\
&+& \left\lb (\pa +b_2 + b_1 )(\pa +b_1 )a_1 \right\rb
P_{s-3}^{(3)}(b_3 ,b_3 +b_2 , b_3 + b_2 + b_1) \; ,\qquad (s\geq 3) \lab{4-6}
\er
where again we used the multiple \faa polynomials \rf{multifaa}.

\lskip
{\large {\bf 5. Outlook and Discussion}}
\lskip
The strategy of this paper was to start with the two-boson KP system
and work out
the higher-boson  representations by a recurrence procedure.
In this way we achieved description of multi-boson KP hierarchies
through abelianization of the first Poisson structure.
The simplicity of the final result rises hopes for future
applications of our method.
Let us briefly indicate possible directions of future investigations.
The corner stone of our construction, the two-boson KP hierarchy,
has recently been a
subject of a quantization attempt \ct{west} promoting the
classical relation (in fact, gauge equivalence)
with the non-linear Schr\"{o}dinger (NLS) hierarchy
to the quantum case.
It seems natural to expect that one can extend this quantization
procedure to four-, six-, etc. boson systems adding successively
quantum NLS hierarchies according to our recurrence relation.

Another question, naturally arising  from our analysis,
is whether the result we obtained
could be used  to gain a new understanding of the lattice hierarchies
connected with
the multi-matrix models. In parallel to our observation  in
this paper one  could expect
a convenient  redefinition  of  fields in the lattice hierarchies
similar to the abelianized
representation of the  pseudo-differential multi-boson KP
Lax operators.

We point out that  the abelianization  construction
(Darboux coordinates)  works so far
only in the context of  the first KP Hamiltonian structure.
In view of the existence
of a compatible  second bracket structure in the unconstrained KP
hierarchy, it will be interesting to  study how the
latter is  affected by  the Poisson reduction and what possible
form the abelianization will take in this framework.
\lskip
{\bf Acknowledgments.} H. A. thanks the Physics Department for hospitality at
Ben-Gurion University of the Negev.

\end{document}